\documentclass[twocolumn]{aastex631}

\usepackage{orcidlink} 
\usepackage[T1]{fontenc}
\usepackage{ae,aecompl}

\usepackage{graphicx}	
\usepackage{amsmath}	
\usepackage{amssymb}	
\usepackage{float}
\usepackage{natbib}

\usepackage{hyperref}
\hypersetup{
	colorlinks=true,
	urlcolor=blue,    
	linkcolor=black,  
	citecolor=blue,   
}

\defcitealias{Pittordis_2018}{PS18}
\defcitealias{Pittordis_2019}{PS19}
\defcitealias{Badry_2021}{ERH}
\defcitealias{Banik_2018_Centauri}{BZ18}

\hypersetup{citecolor=blue, 
            linkcolor=red, 
            menucolor=blue, 
            urlcolor=blue}  

\newcommand{\cm}{{~\rm cm}}
\newcommand{\m}{{~\rm m}}
\newcommand{\km}{{~\rm km}}
\newcommand{\s}{{~\rm s}}
\newcommand{\h}{{~\rm h}}

\newcommand{\g}{{~\rm g}}
\newcommand{\G}{{~\rm G}}
\newcommand{\K}{{~\rm K}}

\newcommand{\yr}{{~\rm yr}}

\begin{document}

\title{High mass accretion rates onto evolved stripped-envelope massive stars by jet-induced mass removal}


\author{Yotham Cohen\,\orcidlink{0009-0001-5629-4381}}\author{Ealeal Bear\,\orcidlink{0000-0002-3592-1526}}\author{Noam Soker\,\orcidlink{0000-0003-0375-8987}}
\affiliation{Department of Physics, Technion, Haifa, 3200003, Israel; ealeal44@technion.ac.il; soker@technion.ac.il}

\begin{abstract}
Simulating one-dimensional stellar evolution models with \textsc{mesa}, we show that removing the outer inflated envelope of a mass-accreting evolved stripped-envelope star, like a Wolf-Rayet (WR) star, substantially moderates the stellar expansion during accretion at high-mass accretion rates. We study the accretion onto a star via an accretion disk, which launches jets that remove the high-entropy outer layers of the inflated envelope. This is the `jetted mass removal accretion scenario.' By manually removing the entire hydrogen-rich envelope from a red supergiant, we build a hydrogen-deficient WR stellar model with a mass of $6.03 M_\odot$ and a radius of $0.67 R_\odot$. We then accrete mass onto it at a high rate. We mimic the real process of simultaneous mass addition near the equatorial plane and jet-induced mass removal from the outer envelope by dividing the accretion period into hundreds of pulses: in the first half of each pulse, we add mass, and in the second, we remove a fraction of this mass. The removal of tens of percent from the added mass decreases the stellar expansion by a factor of $\simeq 2-5$.  Our results show that WR stars can maintain a deep potential well and not expand much while accreting mass at high rates. This allows the formation of an accretion disk and the liberation of large amounts of gravitational energy. Our results strengthen models of intermediate-luminosity optical transients, such as luminous red novae, in which a non-degenerate star accretes at high rates and launches jets that power the transient event. 
\end{abstract}


\section{Introduction}
\label{sec:intro}

When the accreted mass onto a compact object has sufficiently high specific angular momentum, the accretion close to the compact object is via an accretion disk or a belt. In many astrophysical objects, these accretion disks launch a bipolar outflow, namely, two opposite jets (we refer to jets even when the half-opening angle is very large, close to $90^\circ$). 
The jets interact with their environment, including the accreting object and its immediate vicinity, and the large-scale reservoir of the accreted mass. The interaction involves feedback processes, which belong to one of two classes: the negative feedback cycle and the positive feedback cycle. 

In the negative component of the feedback mechanism, the jets heat the mass reservoir and/or expel mass from it, thereby decreasing the mass accretion rate and their power. 
The largest and most energetic class of systems with a negative jet-feedback cycle is cooling-flow clusters of galaxies, in which the cooling of hot gas feeds the central supermassive black hole with cold gas. The jets inflate bubbles that heat the intracluster medium, reducing its radiative cooling and, consequently, lowering the accretion of cold gas, which in turn reduces the jets' power (e.g., reviews by \citealt{McNamaraNulsen2012, Fabian2012, Soker2016Rev}). Similar feedback might occur during galaxy formation and evolution (e.g., \citealt{Soker2016Rev, HardcastleCroston2020}). 

In other environments, the jets drive negative feedback by inflating the mass reservoir or expelling mass from it, both of which reduce the density, thereby decreasing the mass accretion rate and, consequently, the jets' power (see \citealt{Soker2016Rev} for a review comparing the negative jet feedback mechanism in different astrophysical types of objects).  In common envelope evolution (CEE), where the compact companion accretes mass from the envelope of a giant star and launches jets, the jets deposit energy into the giant envelope, inflate it (e.g., \citealt{MorenoMendezetal2017, Chamandyetal2018, Shiberetal2019, LopezCamaraetal2020, LopezCamaraetal2022, ZouChamandyetal2022}), and remove some mass (e.g., \citealt{LopezCamaraetal2019, Schreieretal2025}). The inflation and mass removal reduce the envelope density in the compact object environment, thereby affecting the jets' power; for one-dimensional (1D) simulations of the CEE jet negative feedback mechanism, see \cite{Gricheneretal2021} and \cite{WeinerSoker2025}. For 3D simulations, see, e.g., \cite{LopezCamaraetal2019} and \cite{Hilleletal2022}. During grazing envelope evolution, the jets the companion launches as it grazes the giant's envelope remove most of the mass from the envelope's outskirts, affecting both the jets' power and the orbital motion in a negative feedback cycle (e.g., \citealt{Soker2015GEE, Shiber2018, ShiberIaconi2024}). 
In the jittering jets explosion mechanism of core-collapse supernovae, the jets launched by the newly born neutron star explode the massive star (e.g., \citealt{PapishSoker2011, Soker2025Learning}). The negative feedback cycle is short and terminal, namely, within a few seconds, the jets explode the star, which is the mass reservoir, and completely shut themselves off.  

Alongside the negative feedback processes, there are positive feedback processes where the jets facilitate mass accretion through an accretion disk, thereby maintaining their power and preventing a shutoff. The jets launched by the accretion disk remove angular momentum and energy from the immediate surroundings of the mass-accreting body, allowing more gas to flow in and continue the accretion process at a higher rate than without these jet effects. The removal of energy from the vicinity of accreting objects by jets is crucial to allow for the formation of powerful jets (e.g., \citealt{Shiberetal2016, Chamandyetal2018}).

We (\citealt{BearSoker2025, ScolnicBearSoker2025}) recently proposed an additional type of positive feedback process where the jets operate on the mass-accreting star rather than on the accretion process onto the star. As a main-sequence star accretes mass at a high rate, it expands   (e.g., \citealt{SchurmannLanger2024, MukhijaKashi2025a, MukhijaKashi2025b, MukhijaKashi2026a}),  making its potential well shallower, hence reducing the jets' power. The star might even expand to the degree that it prevents the formation of the accretion disk. In these two studies, we considered a process in which jets launched by the accretion disk remove the outermost layers of the mass-accreting star's envelope. We mimic the jet-induced mass removal by removing the outer layer of 1D stellar models with the stellar evolutionary code \textsc{mesa}. We demonstrated that efficient mass removal of the outer, high-entropy layers allows for a high mass accretion rate while substantially reducing stellar expansion. In this study, we perform similar simulations, but for evolved stripped-envelope massive stars.   

There are strong observational motivations for studying this positive feedback cycle through mass removal. These are the observations that show the ejecta of luminous red novae and other intermediate luminosity optical transients (ILOTs) to be bipolar (e.g., \citealt{Kaminski2024, ZainMobeenKaminski2025}). Bipolar morphologies support powering and shaping by jets (e.g.,  \citealt{Soker2023BrightILOT, Soker2024, ZainMobeenKaminski2025}). Although other powering processes exist, like ejecta recombination and the collision of fresh ejected with earlier ejecta in and near the equatorial plane (e.g., \citealt{Pejchaetal2016a, Pejchaetal2016b, Pejchaetal2017, MetzgerPejcha2017, HubovaPejcha2019}), we consider mass accretion onto the compact companion via an accretion disk and the launching of jets (e.g., \citealt{Soker2020ILOTjets, SokerKaplan2021RAA}). For a powerful transient, the accreting non-degenerate star should not expand significantly as it accretes mass, to maintain a deep potential well. The study of these transients has been a hot subject for decades (e.g., \citealt{Mouldetal1990, Bondetal2003, Rau2007, Ofek2008, Masonetal2010, Kasliwal2011, Tylendaetal2013, Kasliwaletal2012, Kaminskietal2018, BoianGroh2019, Caietal2019, Kashietal2019Galax, Blagorodnovaetal2020, Banerjeeetal2020, Howittetal2020, Jones2020, Kaminskietal2020Nova1670, Kaminskietal2021Nova1670, Klenckietal2021, Stritzingeretal2020AT2014ej, Stritzingeretal2020SNhunt120, Blagorodnovaetal2021, Mobeenetal2021, Pastorelloetal2021, Pastorelloetal2023, Addisonetal2022, Caietal2022,  Wadhwaetal2022, Kaminskietal2023, Karambelkaretal2023,  ZainMobeenetal2024, Kaminski2024, HatfullIvanova2025, Kirilovetal2025, Reguittietal2025, Traninetal2025, Valerinetal2025a, Valerinetal2025b}). 
Our study aims to demonstrate that evolved stripped-envelope massive stars can accrete mass at a high rate while launching jets, a process that may power ILOTs.\footnote{Here ILOTs are transients powered by gravitational energy (for earlier use of this term see, e.g., \citealt{Berger2009, KashiSoker2016Terms, MuthukrishnaetalM2019}), and luminous red novae are ILOTs where the two stars merge to leave one remnant \citep{KashiSoker2016Terms}. Other studies use some different definitions (e.g., \citealt{Jencsonetal2019, Caietal2022b, PastorelloMasonetal2019, PastorelloFraser2019}). } 
The theoretical motivation for studying high-mass accretion rates is the role that jets may play in CEE and in grazing-envelope evolution, as discussed above.

Most studies of strong binary interaction with non-degenerate stars, like ILOTs and CEE, consider at least one star to be a main-sequence star (e.g., \citealt{Tylendaetal2011, Ivanovaetal2013a, Nandezetal2014, Kaminskietal2015, Pejchaetal2016a, Pejchaetal2016b, Soker2016GEE, Blagorodnovaetal2017, MacLeodetal2017, MacLeodetal2018,  Segevetal2019, Howittetal2020, MacLeodLoeb2020, Qianetal2020, Schrderetal2020, Blagorodnovaetal2021, Addisonetal2022, Zhuetal2023, Tylendaetal2024}), with some considering also sub-stellar companions (e.g., \citealt{RetterMarom2003,  Metzgeretal2012, Yamazakietal2017, Kashietal2019Galax, Gurevichetal2022, Deetal2023, Oconnoretal2023}).
In this study, we will consider evolved striped-envelope stars, including Wolf-Rayet (WR) stars. Such companions might be relevant to the major outbursts of some luminous blue variables, which are thought to be powered by accretion (e.g., \citealt{MukhijaKashi2026b}). 
We describe our numerical procedure to mimic the jet-induced mass removal in striped-envelope stars in Section \ref{sec:Method}, and our results in Section \ref{sec:Results}. We summarize in Section \ref{sec:Summary}.  
 
\section{Numerical method}
\label{sec:Method}

\subsection{The numerical code}
\label{subsec:NumericalCode}
We use version 24.03.1 of the stellar evolution code Modules for Experiments in Stellar Astrophysics (\textsc{mesa}; \citealt{Paxtonetal2011, Paxtonetal2013, Paxtonetal2015, Paxtonetal2018, Paxtonetal2019, Jermynetal2023}) in its single star mode. Our simulation, building the evolved star is based on the example of ( \textit{$20M\_pre\_ms\_to\_core\_collapse$}). 
Our changes to the example are:
${\text{thermoline\_coef=2}}$ was changed from 1 based on the example of MESA 23.5.1 to help with convergence. Furthermore, the following parameters were set to 
$\Delta{\textit{\_XHe\_cntr\_hard\_limit}}= 0.02$,
$\Delta{\textit{\_XC\_cntr\_hard\_limit}}= 0.02$,
$\Delta{\textit{\_XNe\_cntr\_hard\_limit}}= 0.02$,
$\Delta{\textit{\_XO\_cntr\_hard\_limit}}= 0.02$ ,
$\Delta{\textit{\_XSi\_cntr\_hard\_limit}}= 0.02$  as in the example of MESA 23.5.1,  and in our previous simulations. This value ensures the simulations converge.

\subsection{Building the WR star}
\label{subsec:WRstar}
We divide our simulations into three main stages: A1, A2, and A3. 
\textit{Stage A1.} In Stage A1, we evolve a zero-age main sequence (ZAMS) star of mass $M_{ZAMS}=20M_\odot$. It experiences helium core burning, and a carbon core appears at $ 8.2 \times 10^6 \yr$ when the stellar mass is $M_{A1}=19.671M_\odot$. The wind mass-loss is `Dutch' with a scaling factor of 0.8 \citep{MaederMeynet2001}.   We terminate Stage A1 when the red supergiant radius is very large, and there is already a carbon and oxygen core, because we assume that at that stage, it engulfs a companion star that, via a common envelope evolution, removes the hydrogen-rich envelope (see also Section \ref{sec:Summary}). The exact value of the radius is of no significance, so we terminate stage A1 at a round evolution time of $t_{\rm f, A1}=9\times 10^6\yr$. 
At this time, the radius was $R_{\rm f, A1}=773.2R_\odot$, the carbon core was $M_{\rm C}=2.12 M_\odot$, and the stellar mass was $M_{\rm f, A1}=17.155 M_\odot$. We list the parameters characterizing Stage A1 in the third row of Table \ref{Tab:Table1}. 
\begin{table*}[]
\centering
\begin{tabular}{|l|l|l|l|l|l|l|l|l|l|l|}
\hline
stage & $t_{\rm f}$ & $\dot M_{\rm ex}$   & $R_{\rm i}$     & $R_{\rm f}$     & $M_{\rm i}$     & $M_{\rm f}$  &$M_{\rm H,f} $ &$M_{\rm {He,f}}$ & $M_{\rm {C12,f}} $ &$M_{\rm {O16,f}}$ \\ \hline
      & yr         & $M_\odot \yr^{-1}$  & $R_\odot$ & $R_\odot$ & $M_\odot$ & $M_\odot$ & $M_\odot$ & $M_\odot$ & $M_\odot$ & $M_\odot$ \\ \hline
A1    & $9 \times 10^6$   & 0        & 5.18      & 773.20    & 19.99     & 17.16  & 6.91 & 6.58 & 2.12 & 1.36   \\ \hline
A2    & $9.001112 \times 10^6$   & -0.01    & 773.14    & 0.67      & 17.15     & 6.025599   & $8.58 \times 10^{-5}$ & 2.51 & 2.1 & 1.31   \\ \hline
\end{tabular}
\caption{The first two stages, which are common to all simulations of Stage A3. Stage A1 follows the star from the ZAMS until the formation of a red supergiant. 
In Stage A2, we mimic a strong binary interaction and manually remove the hydrogen-rich envelope and part of the Helium envelope to transform the star into a stripped-envelope star (a WR star). The columns in the table are as follows: $t_{\rm f}$ is the stellar age at the termination of the stage; $\dot M_{\rm ex}$ is the artificial (or external) mass removal rate in Stage A2 (much above the regular wind mass loss rate); $R_{\rm i}$ and $R_{\rm f}$ are the initial and final radius of each stage, respectively; $M_{\rm i}$ and $M_{\rm f}$ are the initial and final mass of each stage, respectively; $M_{\rm H,f}$, $M_{\rm He,f}$, $M_{\rm C,f}$, and $M_{\rm O,f}$ are the final total mass of hydrogen, helium, carbon, and oxygen respectively.
}
\label{Tab:Table1}
\end{table*}

\textit{Stage A2.} It begins when stage A1 ends (i.e., when the core is transforming into carbon). We start this stage at $t_{\rm f, A1}=9 \times 10^6\yr$. The star is a red supergiant, and we mimic a strong binary interaction, e.g., a common-envelope evolution, and manually remove the hydrogen envelope and most of the helium. The amounts of removed hydrogen and helium are $6.9089M_\odot$ and $4.066M_\odot$, respectively. The hydrogen is completely removed (the total amount of hydrogen that we have at the end of stage A2 is $M_{\rm f, A2, H}=8.58 \times 10^{-5} M_\odot$), while the helium mass that remains is $M_{\rm f, A2, He}=2.51M_\odot$. At the end of Stage A2, the stellar radius and mass are $R_{\rm f, A2}=0.67R_\odot$ and $M_{\rm f, A2}=6.025M_\odot$ respectively. The duration of Stage A2 is $1112 \yr$. We list the parameters characterizing Stage A2 in the fourth row of Table \ref{Tab:Table1}. 

\begin{figure}[]
\centering
\includegraphics[trim=0cm 0cm 0cm 0cm, clip, width=\linewidth]{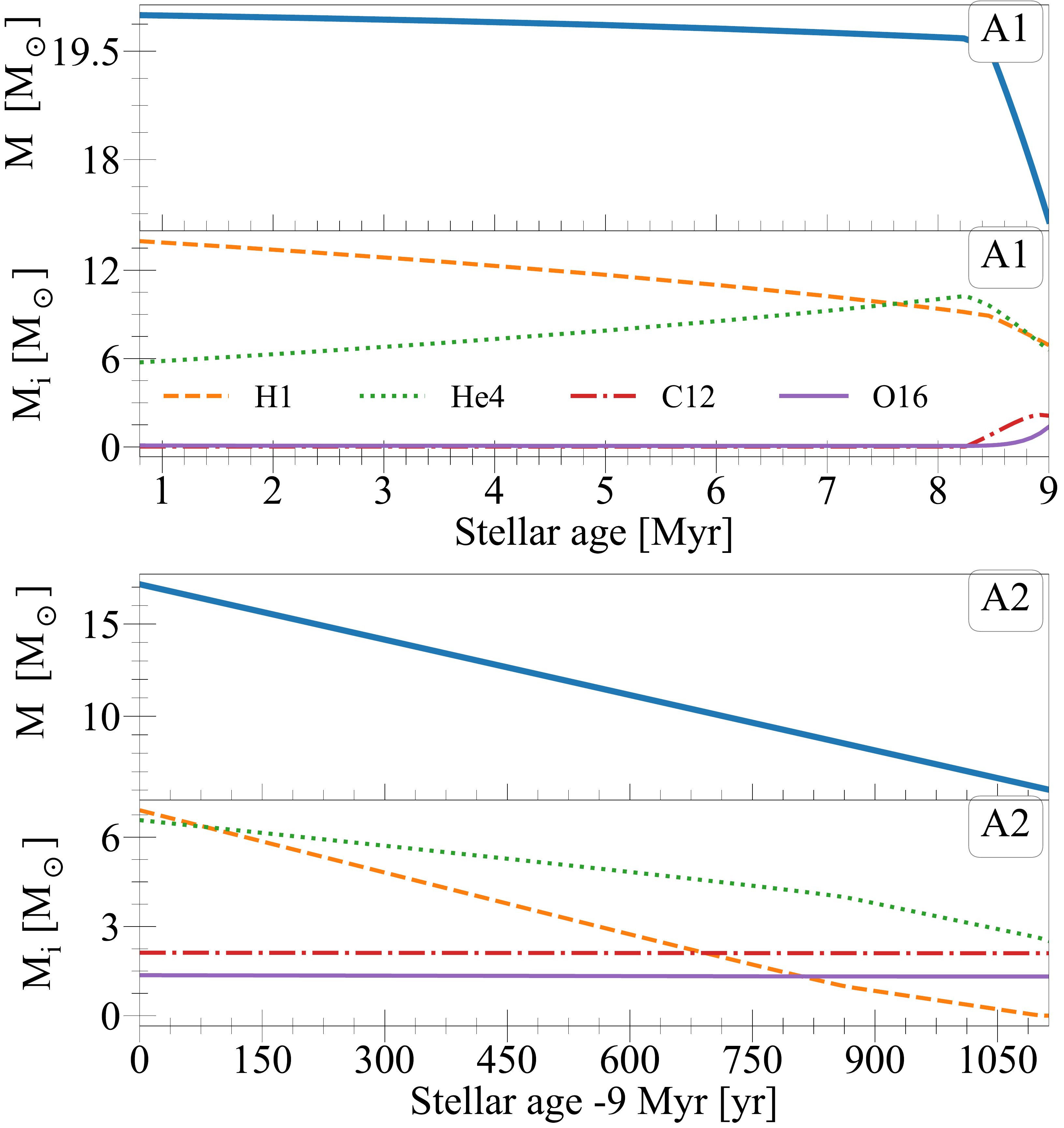}
\caption{The stellar mass (solid thick blue line), $M$, and masses of four isotopes, $M_{\rm i}$ as indicated, as a function of time. The upper two panels correspond to Stage A1, and the lower two panels correspond to Stage A2. 
The mass loss in Stage A1 is the regular stellar wind. In contrast, in Stage A2, we manually set a much higher mass-loss rate to mimic the external effects of a binary companion, e.g., common-envelope evolution. In the upper panels, the time is the stellar age in Myr; in the lower panels, it is the stellar age minus 9 Myr. Stage A2 ends in a stripped-envelope star (a WR star); see Table \ref{Tab:Table1}.  
}
\label{fig:A1_A2_mass_time}
\end{figure}

In Figure \ref{fig:A1_A2_mass_time}, we present the total stellar mass and masses of four isotopes during Stage A1 in the upper panels and during Stage A2 in the lower panels. Note that the time scale in the upper panel is in Myr, whereas in the lower panel it is the age minus 9 Myr. 

\subsection{The mass accretion scheme}
\label{subsec:MassAccretionScheme}
\textit{Stage A3. } Stage A3 contains multiple different simulations with different parameters, all starting with $R_{\rm f, A2}=0.67R_\odot$ and $M_{\rm f, A2}=6.026M_\odot$. We follow the prescription of \cite{ScolnicBearSoker2025} for mass addition and removal in pulses. Each pulse consists of mass accretion followed by mass removal.  The procedure of mass addition and removal mimics the mechanism by which jets launched by the accretion disk remove mass from the outskirts of the star. The accretion is through the accretion disk in and near the stellar equator. Our pulses are not determined by the radius as in \cite{ScolnicBearSoker2025} but in a different way. 
 
In each pulse, the duration of the removal phase is equal to that of the addition phase. Simulations' B', 'C', and 'D' differ in their net accretion. We do not deposit energy into the accretion energy to the envelope, as we assume that the accretion is via an accretion disk and that jets carry most of the accretion energy, either directly or by removing the high-entropy outer envelope zones (see \citealt{BearSoker2025, ScolnicBearSoker2025}).
To find out how to divide the pulses, we simulate cases with constant accretion rates, only mass addition (no pulses), until the star reaches $10R_\odot$ (cases B, C, and D), or for a prescribed duration ($0.1 \yr$ in case E and $0.984 \yr$ in case F). Each of these simulations is marked with a subscript '0' in Table \ref{Tab:Table2} and in the relevant figures. Once we have the total time duration for each case, we divide it into $N_p$ equal-time pulses. During half of the pulse duration, we add mass, and then, for the other half, we remove mass at a lower rate. We simulated cases with different values of $N_p$; case '0' has no pulses, only mass addition; all the others have 50 or more pulses. Simulations 'E' and 'F' are the same as the other simulations in Stage A3, except they are done at a high rate of accretion where $\Delta t= 0.1\yr$ and $\Delta t= 0.98399\yr$ for 'E' and 'F' respectively,  and we stop at this time regardless of the radius.  
We summarize the different Stage A3 cases in Table \ref{Tab:Table2}. 
\begin{table*}[]
\centering
\begin{tabular}{|l|l|l|l|l|l|l|l|l|l|l|l|l|}
\hline
stage & $\Delta t$ & $\dot M_{\rm acc}$ & $N_p$ & $\dot M_{\rm rm}$ & $\dot M_{\rm add}$ & $\alpha$ & $R_{\rm f}$ & $M_{\rm f}$  \\ \hline
 & $\yr$ & $M_\odot \yr^{-1}$  &    & $M_\odot \yr^{-1}$  & $M_\odot \yr^{-1}$  & $R_\odot$ & $M_\odot$ & $M_\odot$ \\ \hline
B0 & 0.785357 & 0.015 & 0 & 0 & 0.015 & NA & 10.00 & 6.03738  \\ \hline
B1 & 0.785357 & 0.015 & 100 & 0.000303 & 0.030303 & 0.01 & 7.72 &  6.03738  \\ \hline
B2 & 0.78536 & 0.015 & 100 & 0.0075 & 0.0375 & 0.20 & 7.55 & 6.03738  \\ \hline
B3 & 0.78536 & 0.015 & 100 & 0.015 & 0.045 & 0.33 & 7.25 &  6.03738  \\ \hline
B4 & 0.78536 & 0.015 & 100 & 0.060 & 0.090 & 0.67 & 5.45 & 6.03738  \\ \hline
B41 & 0.78536 & 0.015 & 200 & 0.060 & 0.090 & 0.67 & 5.67 & 6.03738  \\ \hline
B42 & 0.78536 & 0.015 & 50 & 0.060 & 0.090 & 0.67 & 5.41 &  6.03738  \\ \hline
B5 & 0.78536 & 0.015 & 100 & 0.120 & 0.150 & 0.80 & 4.17 &  6.03738  \\ \hline
B51 & 0.78536 & 0.015 & 200 & 0.120 & 0.150 & 0.80 & 4.20 & 6.03738  \\ \hline
B52 & 0.78536 & 0.015 & 50 & 0.120 & 0.150 & 0.80 & 4.11 &  6.03738  \\ \hline
B6 & 0.78536 & 0.015 & 100 & 0.270 & 0.300 & 0.90 & 2.76 &  6.03738  \\ \hline
B61 & 0.78536 & 0.015 & 200 & 0.270 & 0.300 & 0.90 & 2.96 &  6.03738  \\ \hline
B62 & 0.78536 & 0.015 & 50 & 0.270 & 0.300 & 0.90 & 2.54 & 6.03738  \\ \hline
C0 & 0.25264 & 0.05 & 0 & 0 & 0.05 &  NA & 10.00 & 6.03823  \\ \hline
C1 & 0.25264 & 0.05 & 100 & 0.9 & 1 & 0.90 & 1.96 & 6.03823  \\ \hline
D0 &0.1968 & 0.07 & 0 & 0 & 0.50 & 0.00 & 10.00 & 6.03937  \\ \hline
D1 & 0.1968 & 0.07 & 100 & 1.260 & 1.400 & 0.90 & 1.94 & 6.03937  \\ \hline \hline
E0 & 0.1 & 0.07 & 0 & 0 & 0.07 &  NA & 4.56 & 6.03260  \\ \hline
E1 & 0.1 & 0.07 & 50 & 0.21 & 0.35 & 0.60 & 2.08 & 6.03260 \\ \hline
E2 & 0.1 & 0.07 & 50 & 0.560 & 0.700 & 0.80 & 1.56 & 6.03260  \\ \hline
E3 & 0.0984 & 0.07 & 50 & 1.260 & 1.400 & 0.90 & 1.25 & 6.03249  \\ \hline
F0 & 0.98399 & 0.07 & 0 & 0 & 0.07 &  NA & 45.60 & 6.09448  \\ \hline
F1 & 0.98399 & 0.07 & 500 & 0.560 & 0.700 & 0.80 & 17.22 & 6.09448  \\ \hline
F2 & 0.98399 & 0.07 & 500 & 1.260 & 1.400 & 0.90 & 8.31 & 6.09448  \\ \hline
\end{tabular}
\caption{A table summarizing the different simulations of Stage A3 where mass is added and removed through the pulse prescription, beside cases with subscript `0' that have only constant mass addition rate, i.e., no pulses. The different columns represent the following variables: 'Stage' denotes the simulations, where 'B', 'C' and 'D' correspond to simulations differing in the run duration, as indicated by $\Delta t$; $\dot M_{\rm acc} = (\dot M_{\rm add}-\dot M_{\rm rm})/2$ is the net mass accretion rate, and $N_p$ is the number of pulses. Each pulse combines two phases: mass addition at  rate of $\dot M_{\rm add}$ and mass removal at a rate of $\dot M_{\rm rm}$; $\alpha \equiv {\dot M_{\rm rm}}/{\dot M_{\rm add}}$; $R_f$ and $M_f$ are the final radius and mass, respectively. Simulations marked 'E' and 'F' are done for a specific duration of $0.1 \yr$ and $ 0.98399\yr$ respectively.
The stellar luminosity for all simulations is in the range of $L = 7.14 \times 10^4-1.62 \times 10^5 L_\odot$. 
}
\label{Tab:Table2}
\end{table*}

\begin{figure}[]
\centering
\includegraphics[trim=0cm 0cm 0cm 0cm, clip, width=\linewidth]
{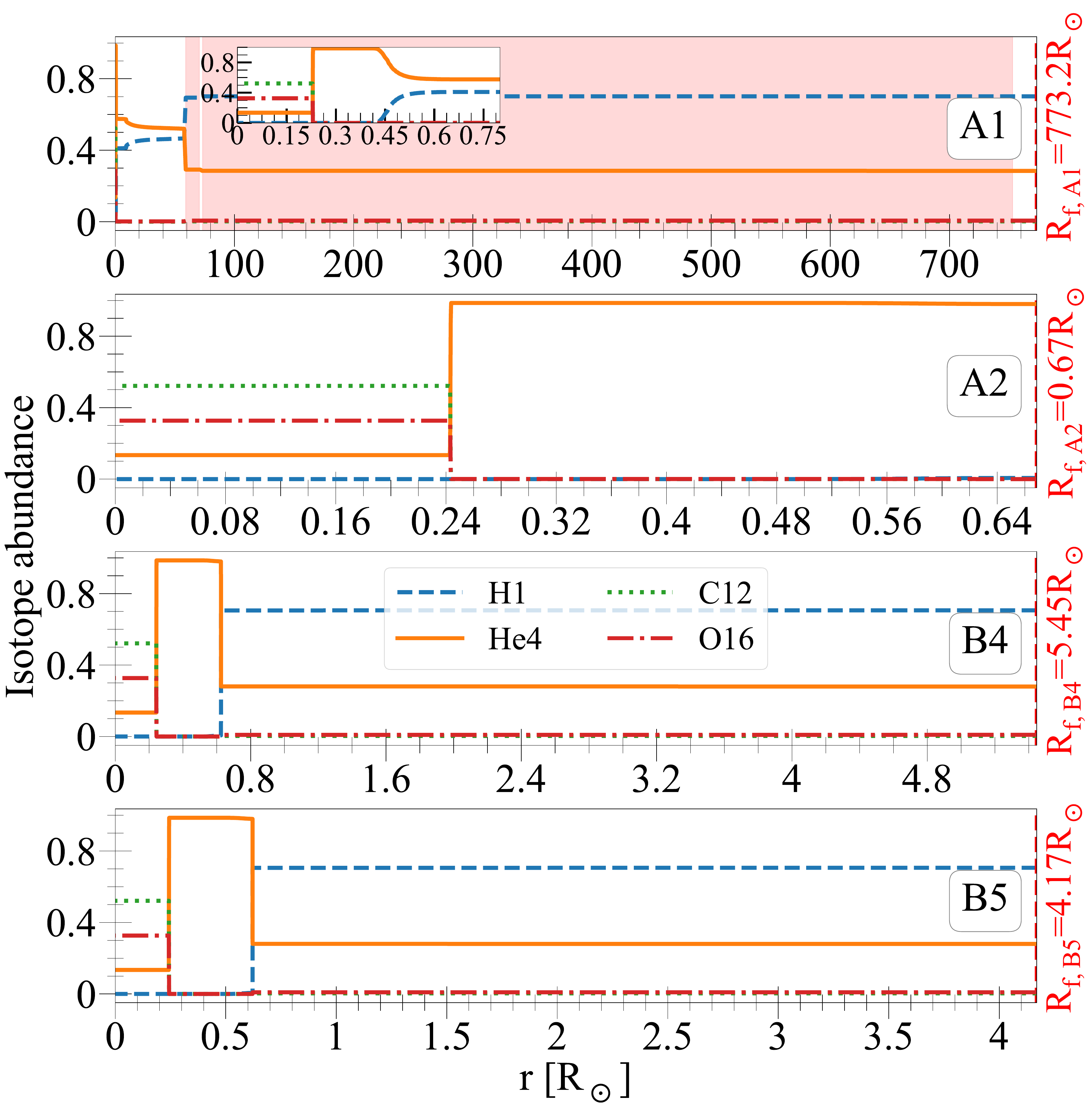}
\caption{Total abundances of four isotopes (see legend) as a function of radius at the end of the three stages (two different simulations of stage A3; see Table \ref{Tab:Table2}). The radius at the end of each phase is to the right of each panel. The pink area in the upper panel marks zones where the convective velocity is $v_{\rm conv}\geq 1 \km \s^{-1}$. The inset in the upper panel zooms in on the core. Note that we accrete hydrogen-rich gas onto the stripped-envelope star (Stage A3: here B4 and B5), which has no hydrogen at the end of Stage A2.     
}     
\label{fig:Mass_element_vs_radius}
\end{figure}

In Figure \ref{fig:Mass_element_vs_radius} we present the abundances of four isotopes in the different stages (two cases of the many in Stage A3). This figure shows the evolution from a hydrogen-rich red supergiant star (end of Stage A1; upper panel) to a stripped-envelope small star (end of Stage A2; second panel). The lower two panels present two different simulations of Stage A3. The envelope at the end of the accretion phase is hydrogen-rich. We accrete hydrogen-rich gas as we assume that the stripped-envelope star accretes from a hydrogen-rich star, like a main sequence star or an evolved star.   

\section{Mass accretion and expansion}
\label{sec:Results}

In simulation B0, we added mass at a constant rate of $0.015 M_\odot \yr^{-1}$ for $0.785357 \yr$, resulting in a net accreted mass of $0.01178 M_\odot$. 
The star expanded from $R_{\rm f, A2}=0.67 R_\odot$ to $R_{\rm f, B0}=10 R_\odot$, at this radius we terminate the simulation. In the other B simulations, we accrete an equal amount of net mass at the same time period, but in pulses during addition-removal cycles. This mimics the removal of the outer envelope layers by the accretion disk's jets. As the column marked $R_{\rm f}$ indicates in Table \ref{Tab:Table2}, this results in a much more moderate stellar expansion. As $\alpha={\dot M_{\rm rm}} / {\dot M_{\rm add}}$ increases, the expansion decreases. However, we cannot increase this too much. For the maximum value we simulated, $\alpha = 0.9$, the removed mass is $0.9$ times the added mass, and the accreted mass, $M_{\rm acc}=M_{\rm add} - M_{\rm rm}$, is $0.1$ times the added mass. If the added mass is accreted from rest at infinity, the terminal velocity (i.e., at large distances) of the ejected mass is given by energy conservation 
\begin{equation}
v_{\rm rm, \infty} =  \frac{v_{\rm esc}}{\sqrt{2}} \sqrt{\frac {M_{\rm acc}}{M_{\rm rm}}} = v_{\rm Kep} \sqrt{\frac {1-\alpha}{\alpha}}  ,
\label{eq:eq1}
\end{equation}
where $v_{\rm esc}$ is the escape velocity from the star, and $v_{\rm Kep}$ is the Keplerian orbital velocity on the surface of the star. We assume that the mass accreted onto the star releases half its binding energy from the virial theorem and that radiation carries a negligible amount of energy during the accretion process. The luminosity of the event results mainly from the collision of the removed mass, namely, the jets, with the circumstellar material farther out  (we elaborate on this process in Section \ref{sec:Summary}).    
For the present stellar model at the beginning of mass addition $v_{\rm Kep}=1310 \km \s^{-1}$, and so for $\alpha=0.9$ we find the terminal velocity of the removed mass to be $v_{\rm rm} =437 \km \s^{-1}$. However, as the star expands, the potential well becomes shallower, and the velocity decreases.  

\begin{figure}[]
\centering
\includegraphics[trim=0cm 0cm 0cm 0cm, clip, width=\linewidth]{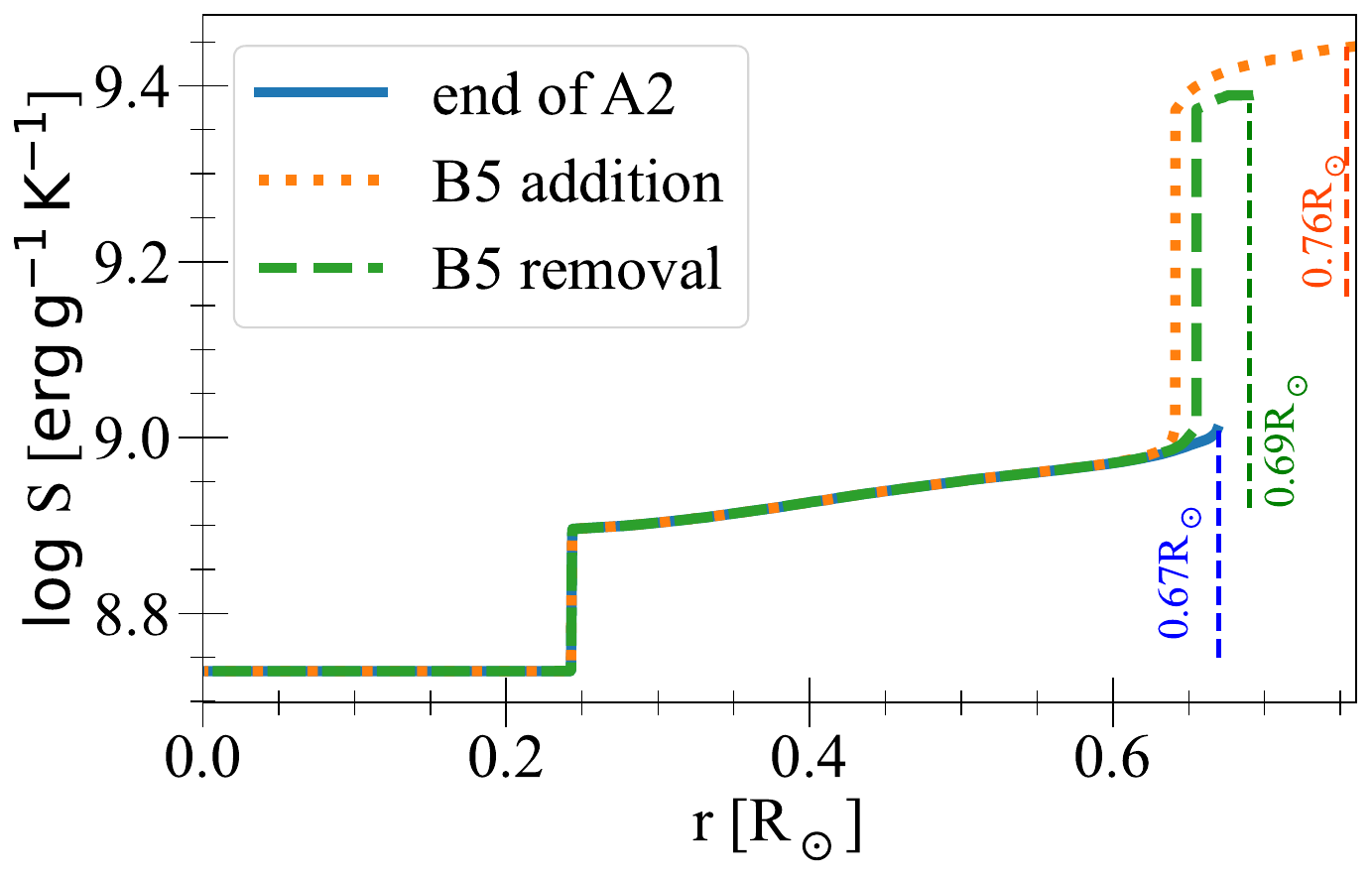}
\caption{Entropy profiles as a function of stellar radius at three times during one pulse of mass accretion in simulation B5. The three profiles overlap in the inner zones. The blue line shows the entropy profile at the end of Stage A2, just before we start the accretion process; only its right-hand, which extends to a radius of $0.67R_\odot$, is visible in this plot. The dotted-orange line shows the entropy profile after mass addition at the end of the first half of the first pulse, and the dashed-green line shows it after mass removal of $80\%$ of the added mass at the end of the pulse. 
}
\label{fig:entropy}
\end{figure}

We mimic accretion via an accretion disk that launches jets. While the star accretes mass from near the equatorial plane, the jets that the accretion disk launches remove high-entropy gas from the envelope outskirts. This has previously been studied in hydrogen-rich stellar models \citep{BearSoker2025, ScolnicBearSoker2025}. The main goal is to prevent large stellar expansion so that the star maintains a deep potential well, so that the accretion process releases a large amount of energy per accreted unit mass. 
Figure \ref{fig:entropy} demonstrates the removal of high entropy gas for simulation B5 (for more details, see the earlier two papers in the series). The solid blue line extending to a radius of $0.67R_\odot$ represents the entropy before mass is added in the first pulse (the profile at the end of stage A2). After the addition of mass, the entropy rises to a large value, and the radius increases to about $0.76 R_\odot$  (orange dotted line). After we remove $80 \%$ of the added mass in the first pulse, the radius decreases to about $0.69 R_\odot$. The numerical code \textsc{mesa} removed the outer part, which here is the higher-entropy zone.    

\begin{figure}[]
\centering
\includegraphics[trim=0.2cm 0.0cm 0.0cm 0.0cm ,clip, scale=0.39]{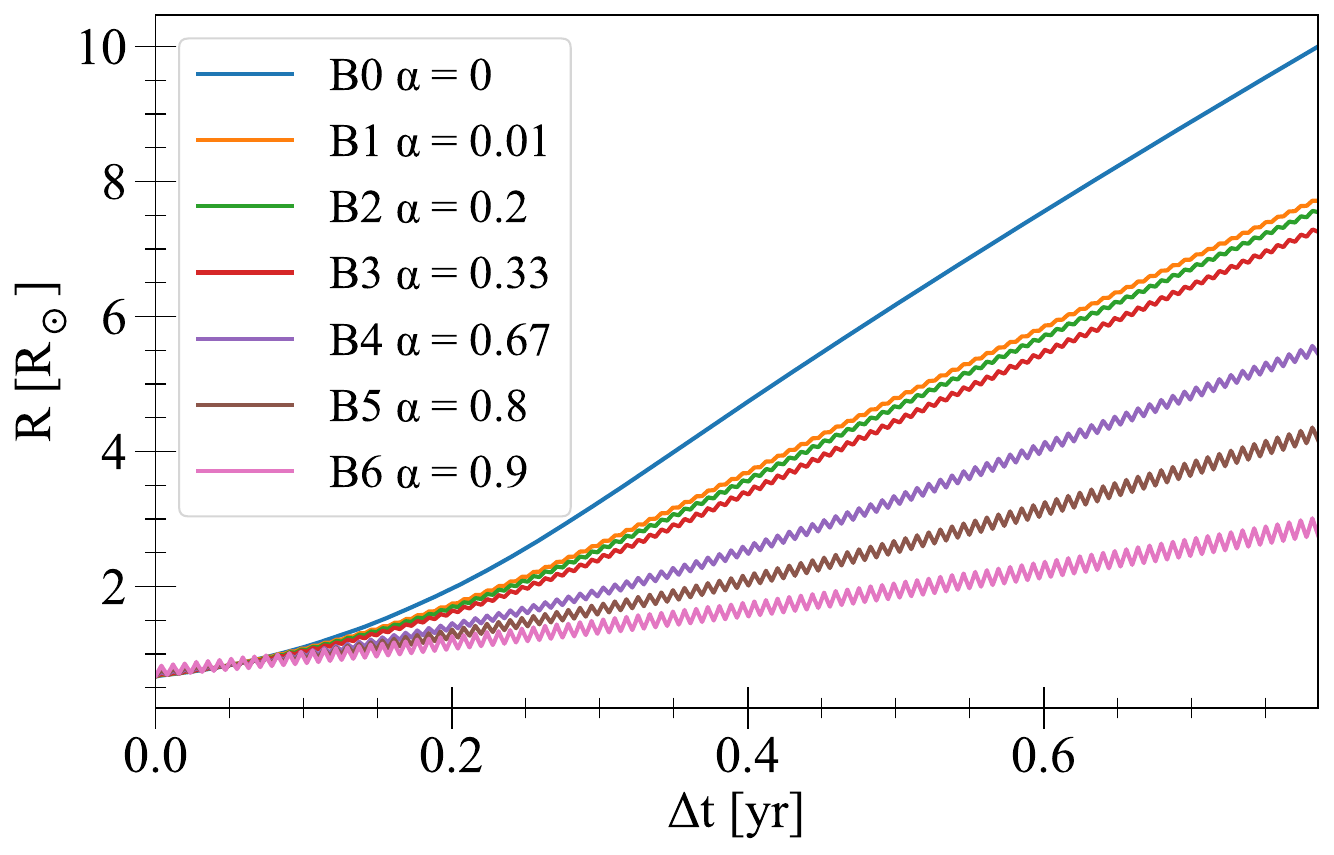}

\caption{ The stellar radius as a function of time measured from the beginning of the accretion process (Stage A3) for simulation group B. In all simulations here the net accretion rate is $\dot M_{\rm acc}=0.015 M_\odot \yr^{-1}$. The upper, smooth blue line shows the expansion as mass is continuously added. The other lines show accretion in pulses: half the pulse time is spent on mass addition, followed by mass removal. $\alpha$ is the ratio of removal to addition rates, i.e., $1-\alpha$ is the fraction of added mass that the star retains. The smaller the fraction of retained mass, the smaller the expansion. } 
\label{fig:alpha}
\end{figure}

\begin{figure}[]
\centering
\includegraphics[trim=0cm 0cm 0cm 0cm, clip, width=0.95\linewidth]{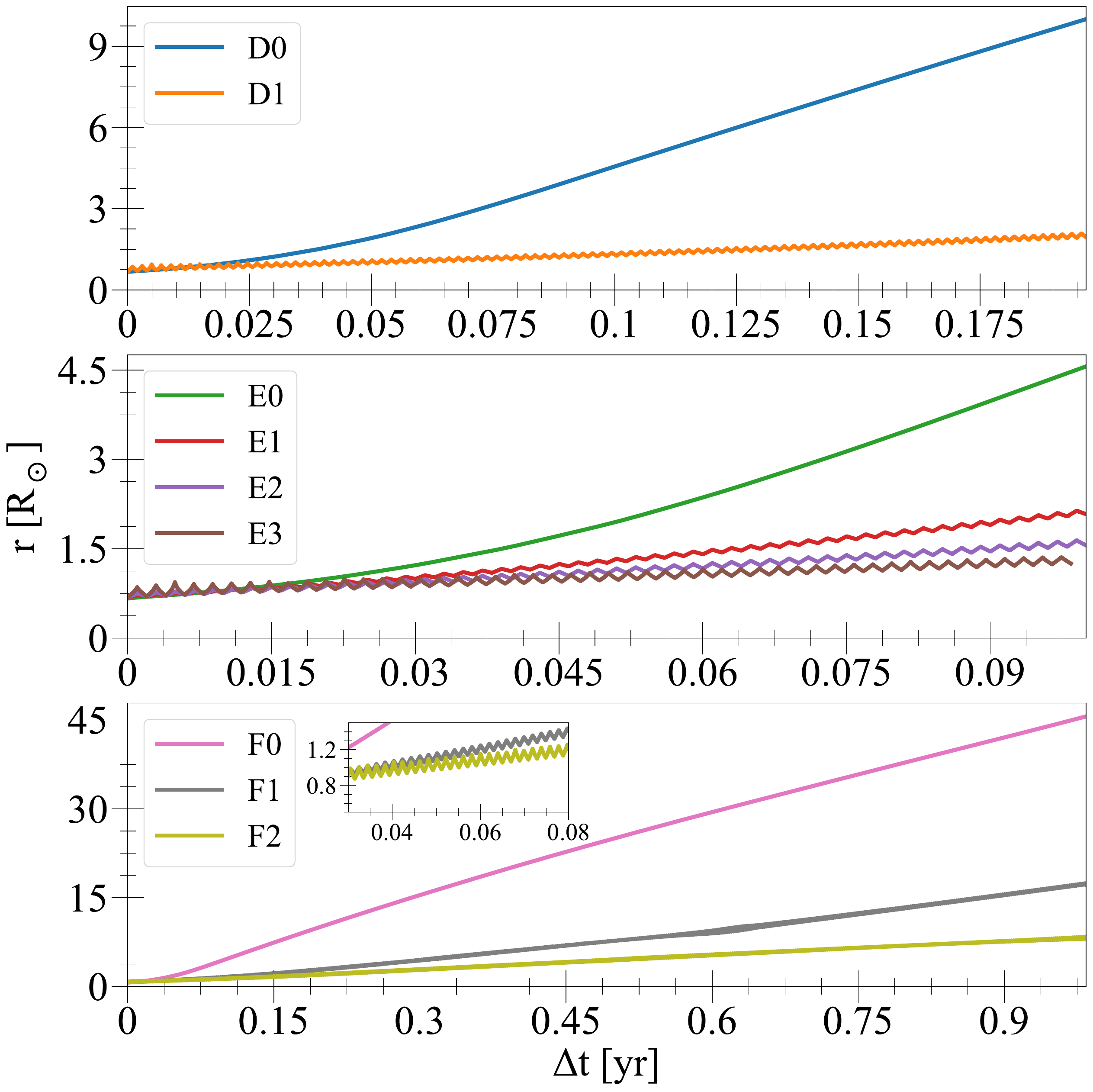}
\caption{Similar to Figure \ref{fig:alpha}, but for cases D, E, and F, which have different mass accretion rates and different durations (see Table \ref{Tab:Table2}). Due to the large number of pulses (500) in simulations F1 and F2, the line width smears the zigzag. In the inset of the lower panel, we expand a short time interval to demonstrate that the zigzag exists.  } 
\label{fig:R_vs_t_SIM_D_E_F}
\end{figure}

Figure \ref{fig:alpha} demonstrates the reduced stellar expansion for several simulations of cases B, all accreting the same amount of mass in the same period of time (Table \ref{Tab:Table2}). This figure shows the evolution of stellar radius over time, and that the larger the value of $\alpha = {\dot M_{\rm rm}}/ {\dot M_{\rm add}}$, the smaller the stellar expansion. In Figure \ref{fig:R_vs_t_SIM_D_E_F} we present the radius as a function of time for cases D, E, and F. These three cases also present the reduced stellar expansion with increasing value of $\alpha$. 

\begin{figure}[]
\centering
\includegraphics[trim=0cm 0cm 0cm 0cm, clip, width=\linewidth]{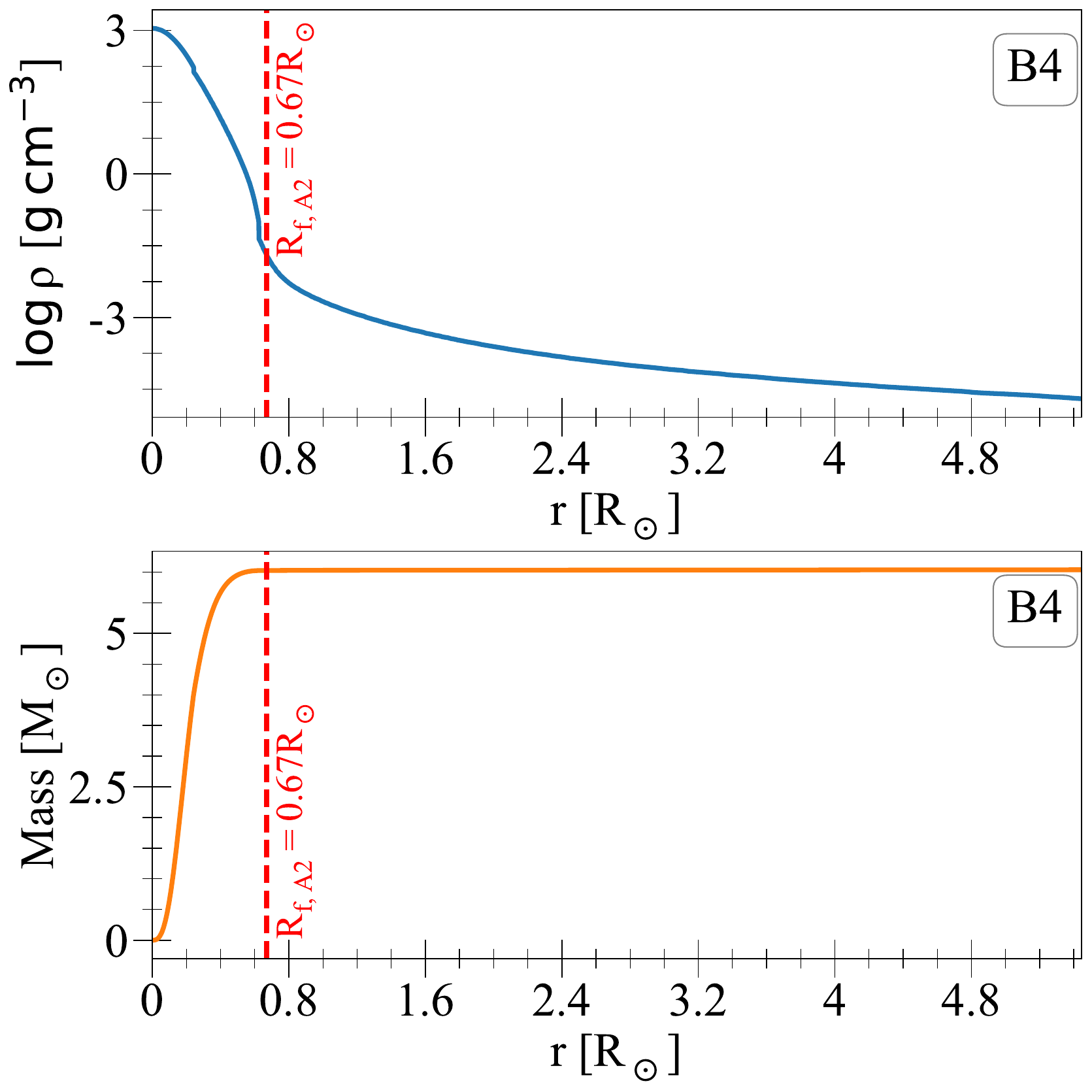}
\caption{Density  (upper panel; log scale) and mass (lower panel) as a function of stellar radius at the end of simulation B4. There are no convection zones. The final radius of Stage A2 (before mass accretion begins) is indicated by a dashed-red line. This figure demonstrates the low density within the inflated envelope due to mass accretion, thereby allowing the accretion disk to penetrate into the outer envelope. }
\label{fig:Density_mass_vs_r_B4}
\end{figure}

Only three-dimensional simulations of the accretion process with a correct treatment of the star can determine the correct value of $\alpha$. These are extremely complicated simulations. We expect the mass removal process to operate in a negative feedback loop. When the star expands more, the outer envelope layers' binding energy decreases, while its volume increases, both of which make mass removal by the jets that the accretion disk launches easier. The condition is that the expanding envelope does not destroy the accretion disk, as \cite{ ScolnicBearSoker2025} have shown is likely to be the case. The reason is that the density of the expanding envelope is lower than the density of the accretion disk \citep{ScolnicBearSoker2025}. In Figure \ref{fig:Density_mass_vs_r_B4} we present the density and mass profiles for simulation B4. These indicate that the expanding envelope has a low mass and low density. This allows the accretion disk to survive within the outer envelope and launch jets that remove the envelope's outer regions. 
 Substituting the accretion rate for simulation B4, $\dot M_{\rm acc} = 0.015 M_\odot \yr^{-1}$, and the stellar mass $M=6 M_\odot$ in equation (2) of \cite{ScolnicBearSoker2025}, we find the density of the accretion disk at the stellar radius $R= 5.45 R_\odot$ at the time we present Figure \ref{fig:Density_mass_vs_r_B4}, to be $\rho_{\rm d} \simeq 3 \times 10^{-6} \g \cm^{-3}$. There is the uncertainty due to the Shakura-Sunyaev viscosity parameter, which is scaled by \cite{ScolnicBearSoker2025} to be $\alpha_{\rm d} = 0.1$. 
This density is about equal to the stellar density on the surface, as seen in Figure \ref{fig:Density_mass_vs_r_B4}. We conclude that the disk can survive in the outer parts of the inflated envelope, even if somewhat distortedted by the stellar envelope.  

\section{Discussion and Summary}
\label{sec:Summary}

This is the third study of the jetted mass removal accretion scenario. The scenario involves accretion onto a non-degenerate star via an accretion disk that launches jets. The jets remove the high-entropy gas in the outskirts of the star, thereby slowing down the stellar expansion. This allows further accretion onto a deep potential well.  
\cite{BearSoker2025} and \cite{ScolnicBearSoker2025} demonstrated this positive jet feedback cycle for massive main-sequence stars, e.g., hydrogen-rich envelope. In this study, we demonstrated this positive feedback cycle for an evolved striped-envelope star with a ZAMS mass of $M_{\rm ZAMS}= 20 M_\odot$, i.e., a WR star.  

 The stellar binary evolutionary scenario we consider is as follows. A massive star, which in our study has a mass of $M_{\rm ZAMS}=20 M_\odot$, engulfs a companion of much lower mass, $M_{\rm com} \simeq 2-5 M_\odot$, as it becomes a red supergiant. The companion spirals in, namely, a common envelope evolution, and removes the entire hydrogen-rich envelope of the primary star. The system is now a close binary of a WR-like star (the stripped-envelope core) and a main-sequence companion. Since the companion accretes mass from the envelope on a timescale much shorter than its thermal timescale (as the common envelope evolution lasts for several years to tens or a few hundred years), its envelope is not thermally relaxed and is inflated. Due to tidal interaction in the post-common envelope evolution between the WR-like star and the companion, they come closer and experience mass transfer of hydrogen-rich material from the companion to the WR-like star. It is this mass transfer that we simulate in this study.  

  In this study, we start the phase of hydrogen-rich mass accretion onto the WR-like star immediately after the removal of its own hydrogen-rich envelope as it evolves to become a red supergiant. However, the envelope of the WR-like star does not change much over a time period of $\simeq 10^5 \yr$. For example, the radius of the WR-like star changes from $R_\ast = 0.668 R_\odot$ at the end of mass removal (end of phase A2, which is the beginning of phase A3 in this study), to  $R_\ast = 0.679 R_\odot$ at $5.2 \times 10^4 \yr$ later and $R_\ast = 0.654 R_\odot$ at $1.13 \times 10^5 \yr$. Our conclusions will not change qualitatively if we start the high-mass accretion rate process any time within $\simeq 10^5 \yr$ from the end of the envelope stripping process.

To mimic this mass accretion process using the one-dimensional numerical code \textsc{mesa}, we conducted simulations that split the accretion into multiple pulses. In each pulse, we add mass in the first half and remove a fraction $\alpha$ of it in the second half. The mass is removed from the outer layers, as the jets are assumed to do. 
Figure \ref{fig:entropy} presents the entropy profiles before we start mass addition, after mass addition, and after mass removal of the first pulse of one simulation. After mass addition, the outer envelope has very high entropy. Mass removal from the outer zones of a high-entropy gas leads to stellar contraction.  
Figures \ref{fig:alpha} and \ref{fig:R_vs_t_SIM_D_E_F} show that this procedure allows accretion of mass that results in a much more moderate stellar expansion than an accretion without mass removal.  

Consider, for example, simulation group E, which lasts for $0.1 \yr$, about 5 weeks. For $\alpha =0.9$ (E3), the final stellar radius is $0.27$ times the radius in the model where we accrete mass at the same rate continuously (E0). This implies a deeper potential well that can release more energy for the same mass accretion rate. The total accreted mass in this one month is $M_{\rm acc}= 0.007 M_\odot$. For an average radius of about $R_\ast  \simeq 1 R_\odot$ for the $\alpha=0.9$ case, the power of the accretion process is  
\begin{equation}
\begin{split}
\dot E_{\rm acc} & \simeq \frac{1}{2} \frac{G M_\ast M_{\rm acc}}{R_\ast} = 6.6 \times 10^6 \left( \frac{M_\ast}{6 M_\odot} \right)
\\ & \times
\left( \frac{\dot M_{\rm acc}}{0.07 M_\odot \yr^{-1}} \right) 
\left( \frac{R_\ast}{1 R_\odot} \right)^{-1} L_\odot .
\label{eq:eq2}
\end{split}
\end{equation}
This power is several tens times the stellar luminosity, which spans the range of $L = 7.14 \times 10^4-1.62 \times 10^5 L_\odot$ in the different simulations of Stage A3 (Table \ref{Tab:Table2}). 
 Due to the optically thick region close to the interacting binary system, most of this energy ends in an outflow, namely, jets, most likely wide jets (or a bipolar disk wind). The collision of this outflow with a circumbinary material converts most of the kinetic energy to thermal energy. The duration of the outburst depends on the duration of the jet-launching process and on the photon diffusion time from the circumbinary material (e.g., \citealt{SokerKaplan2021RAA}).     
If radiation carries $\simeq 10 \%$ of this energy or more, we have a luminous transient (like ILOT or a luminous red novae; see Section \ref{sec:intro} for discussion of these transients and the motivation to study the launching of jets). 
  
 We note the following three properties. (1) We accrete hydrogen-rich material. Therefore, the jets will also be hydrogen-rich. (2) As mentioned above, in the jet-powered ILOT scenario, the jets collide with a slow and more massive circumbinary material, converting most of the jets' kinetic energy to thermal energy, part of which is radiated away (e.g.,  \citealt{SokerKaplan2021RAA}). Therefore, the material that emits the radiation is more massive than the jets; it can be more than an order of magnitude more massive. (3) Most observed ILOTs involve main-sequence stars, rather than stripped-envelope stars. The type of mass-accreting companion we study here is much rarer, and we are not aware of any yet observed ILOT with a WR-like star.

The smaller radius of the star in the jetted mass-removal accretion scenario favors the formation of an accretion disk. A smaller star allows material with lower specific angular momentum to form an accretion disk. This might be significant for accretion from a wind or a giant envelope during common envelope evolution, as in these cases, the angular momentum arises from a small density gradient. 
 
Overall, our study strengthens the jetted mass-removal accretion scenario and, indirectly, the claim that many transient events (ILOTs), such as luminous red novae, can be powered by a non-degenerate star that accretes mass via an accretion disk and launches jets.


\section*{Acknowledgments}

 We thank an anonymous referee for very detailed and useful comments.  An Asher Space Research Institute grant at the Technion supported this research. NS thanks the Charles Wolfson Academic Chair at the Technion for the support.



\label{lastpage}

\end{document}